\documentclass{PoS}
\usepackage{graphicx}
\usepackage{bm}
\usepackage{graphics}
\usepackage{amsmath,amsfonts,amssymb}

\def\Dsl{\hbox{/\kern-.6000em D}} 

\def\dsl{\,\raise.15ex\hbox{/}\mkern-13.5mu D}

\def\ltap{\ \raise.3ex\hbox{$<$\kern-.75em\lower1ex\hbox{$\sim$}}\ }
\def\gtap{\ \raise.3ex\hbox{$>$\kern-.75em\lower1ex\hbox{$\sim$}}\ }
\def\OMIT#1{}

\def\lsim{\mathrel{\raise.3ex\hbox{$<$\kern-.75em\lower1ex\hbox{$\sim$}}}}
\def\gsim{\mathrel{\raise.3ex\hbox{$>$\kern-.75em\lower1ex\hbox{$\sim$}}}}

\def\msb{{\overline{\rm MS}}}

\newcommand{\ord}{{\cal O}}

\def\msb{{\overline{\rm MS}}}

\catcode`\@=11
\def\slash{\mathpalette\make@slash}
\def\make@slash#1#2{\setbox\z@\hbox{$#1#2$}%
  \hbox to 0pt{\hss$#1/$\hss\kern-\wd0}\box0}
\catcode`\@=12 

\newcommand{\be}{\begin{equation}}
\newcommand{\ee}{\end{equation}}
\newcommand{\bea}{\begin{eqnarray}}
\newcommand{\eea}{\end{eqnarray}}

\newcommand{\rot}{\color{red}}
\newcommand{\blau}{\color{blue}}

\definecolor{orange}{rgb}{1,0.5,0}
\definecolor{lila}{rgb}{0.5,0,0.5}
\definecolor{brown}{rgb}{0.6,0.4,0.2}
\newcommand{\orange}{\color{orange}}

\title{Bottom Mass from Nonrelativistic Sum Rules at NNLL\thanks{preprint DESY 13-008, UWThPh-2013-3}}

\ShortTitle{Bottom Mass from Nonrelativistic Sum Rules at NNLL}

\author{\speaker{Maximilian Stahlhofen}\\
        DESY Theory Group, Notkestra\ss e 85, D-22607 Hamburg, Germany\\
        E-mail: \email{maximilian.stahlhofen@desy.de}}

\abstract{
We report on a recent determination of the bottom quark mass from
nonrelativistic (large-$n$) $\Upsilon$ sum
rules with renormalization group improvement (RGI) at next-to-next-to-leading
logarithmic (NNLL) order. The comparison to previous fixed-order analyses shows
that the RGI computed in the vNRQCD framework leads to a substantial stabilization of the theoretical sum rule moments with respect to scale variations. 
A single moment fit ($n=10$) to the available experimental data yields 
$
M_b^{1S}=4.755\pm 0.057_{\rm pert}\pm 0.009_{\alpha_s}
\pm 0.003_{\rm exp}\,\mbox{GeV}
$ 
for the bottom 1S mass and
$
\overline m_b(\overline m_b)= 4.235 \pm 0.055_{\rm pert}
\pm  0.003_{\rm exp} \,\mbox{GeV}
$
for the bottom $\msb$ mass. The quoted uncertainties refer to the perturbative
error and the
uncertainties associated with the strong coupling and the experimental input.
}

\FullConference{Xth Quark Confinement and the Hadron Spectrum,\\
		October 8-12, 2012\\
		TUM Campus Garching, Munich, Germany}

\begin{document}

\section{Introduction}
\label{sectionintro}

Determinations of the bottom quark mass $m_b$ have been the subject of a large number of QCD precision studies in the past. For a summary we refer to Ref.~\cite{Beringer:1900zz}. The bottom mass is an important parameter in numerous theoretical predictions not only within, but also beyond the standard model.

The data from $e^+ e^-$ collisions is the common experimental input in many determinations of $m_b$, because in particular the region close to the $b {\bar b}$ threshold and the $\Upsilon$ resonances of the total cross section are very sensitive to the bottom mass parameter. One classic approach is based on the sum rule~\cite{Novikov:1977dq} that states the equality of the experimental moment
\begin{align}
 P_n^{exp} \,=\,
\int_0^\infty \! \frac{d s}{s^{n+1}}\,R_{b\bar b}(s)
\,,
\label{Pndef1}
\end{align}
where $R_{b\bar b}=\sigma(e^+e^-\to b\bar b+X)/\sigma_{\rm pt}$ is the measured inclusive (hadronic) bottom pair production cross section normalized by $\sigma_{\rm pt}=4\pi \alpha^2/3s$,
and the corresponding theoretical expression $P_n^{th}$ obtained from an operator product expansion (OPE) in QCD.
For not too large $n$ nonperturbative power corrections to $P_n^{th}$ are suppressed and the theoretical prediction is dominated by the perturbative QCD (pQCD) result for an external bottom quark pair. 
Concerning the appropriate theoretical formalism bottom mass determinations from the equation $P_n^{th}(m_b) = P_n^{exp}$ differ depending on the values for $n$.
We distinguish two classes.
 
For $n\lsim 3$ the theoretical moment $P_n^{th}$ is governed by fluctuations at the scale $m_b$. Therefore higher order terms in the OPE typically scale like powers of $\Lambda_{QCD}/m_b$ and the conventional pQCD result is in principle sufficient for a precise determination of the bottom mass. Recent low-$n$ analyses~\cite{Chetyrkin:2009fv,Chetyrkin:2010ic} employ an approximate four-loop pQCD calculation. Ref.~\cite{Bodenstein:2011fv} uses a related variant of this method, where the integration in Eq.~\eqref{Pndef1} is only carried out over a finite range and a compensating term (according to Cauchy's theorem) is added to the theory prediction. Both low-$n$ approaches have the drawback that precise experimental data is currently only available in the region close to the production threshold (and for the $\Upsilon$ resonances) and this deficiency has to be compensated by additional theory input in one way or the other\footnote{In fact the finite energy sum rule used in Ref.~\cite{Bodenstein:2011fv} is equivalent to an infinite energy sum rule, if in the range above the finite energy limit the theory result is used for the total cross section $R_{b\bar b}$ in Eq.~\eqref{Pndef1} and the infinite moment integration converges.}, see e.g. Refs.~\cite{Corcella:2002uu,Chetyrkin:2010ic} for discussions on the corresponding uncertainties.
Power corrections to the additional theory contribution\footnote{related to the fact that the energy integration contour needs to be deformed onto the positive real axis close to the finite energy cutoff} in the moments are commonly assumed to be small.

On the other hand large-$n$ moments, where $4 \lsim n\lsim 10$, receive only negligible contributions from the energy regions beyond threshold and are dominated by the experimentally well-known $\Upsilon$-resonances and hence nonrelativistic bound state dynamics. The corresponding sum rules are therefore often called nonrelativistic or $\Upsilon$ sum rules.
Due to the nonrelativistic nature of the large-$n$ moments in addition to the hard scale $m_b$ the soft scale $m_b/\sqrt{n}$ and the ultrasoft scale $m_b/n$ emerge as relevant short-distance scales and the convergence of the OPE requires the upper limit $n\lsim 10$. 
The hierarchy between these scales induces sizable terms $\propto (\alpha_s \sqrt{n})^k$, the so-called Coulomb singularities, and large logarithms $\propto (\alpha_s \ln(n))^l$ in the perturbative loop expansion.
The resummation of the Coulomb singular terms to all orders can be performed within the effective field theory NRQCD~\cite{Caswell:1985ui,Bodwin:1992ye}.
Extensions of this framework like the pNRQCD~\cite{pNRQCDfirst,Brambilla:1999xf} and the vNRQCD~\cite{Luke:1999kz} formalism also allow the systematic resummation of the logarithmic terms.
The renormalization group improved (RGI) result for the theoretical large-$n$ moments is expressed as a simultaneous expansion in $\alpha_s$ and $1/\sqrt{n}$ and schematically takes the form
\begin{align}
\label{counting}
P_n \,\sim \,\sum_{k,l}(\alpha_s\sqrt{n})^k(\alpha_s\ln(n))^l\,\Big[
1\,\mbox{(LL)};\,
\alpha_s,1/\sqrt{n}\,\mbox{(NLL)};\,
\alpha_s^2,\alpha_s/\sqrt{n},1/n\,\mbox{(NNLL)};\, \ldots
\Big]\,
\end{align}
for the leading logarithmic (LL), next-to leading logarithmic (NLL) and
next-to-next-to leading logarithmic (NNLL) order. 
Prior to the work presented here the RGI bottom mass determination from large-$n$ sum rules reached NLL and partly NNLL level~\cite{Pineda:2006gx,Pineda:2006ri}, but did not include the NNLL running of the dominant heavy quark pair production current. 
Earlier fixed-order analyses~\cite{Voloshin:1995sf,Hoang:1998uv,Melnikov:1998ug,Beneke:1999fe,Hoang:1999ye,Hoang:2000fm} up to next-to-next-to-leading order (NNLO) only resum the Coulomb singularities and count $(\alpha_s\ln(n))^l$ as $\alpha_s^l$ in Eq.~\eqref{counting}. The convergence of the fixed-order expansion however turned out to be rather unsatisfactory, see Ref.~\cite{Battaglia:2003in} for a review. As we will show below RGI computations improve the convergence properties substantially and allow for a reliable and precise determination of the bottom quark mass from nonrelativistic sum rules.

For any determination of the bottom mass parameter with a precision at the percent level it is mandatory to adopt an appropriate short-distance mass-scheme in order to avoid $\ord(\Lambda_{QCD})$ infrared renormalon ambiguities. Suitable mass schemes are the $\msb$ scheme for the low-$n$ sum rules and so-called threshold mass-schemes~\cite{Beneke:1998rk,Hoang:1999zc,Hoang:2000yr,Pineda:2001zq} for the large-$n$ sum rules.

The present talk focuses on the determination of the 1S bottom mass~\cite{Hoang:1999zc} from RGI large-$n$ sum rules and is mostly based on the recently published Ref.~\cite{Hoang:2012us}. In this analysis single moment fits of the mass parameter are carried out including for the first time the almost complete NNLL correction to the theoretical moments. The still missing contribution from the NNLL soft mixing correction to the running of the heavy quark production current can be neglected under the assumption that its size is comparable to the already known soft NNLL terms.

\section{Experimental Moments}

The dominant contribution (87\% - 98\% for $n=6$-$12$) to the experimental moments in Eq.~\eqref{Pndef1} for large $n$ comes from the first four $\Upsilon$ resonances, $\Upsilon(1S)$-$\Upsilon(4S)$, which we construct from their electromagnetic decay widths and masses~\cite{Beringer:1900zz} using the narrow width approximation. 
For the contribution from the threshold region (5.7\% - 1.4\% for $n=6$-$12$) we use BABAR data ~\cite{Aubert:2008ab} in the energy range between $\sqrt{s}=10.62$ and $\sqrt{s}=11.21$ and follow the approach of Ref.~\cite{Dehnadi:2011gc}. 
Finally the continuum region above 11.21 GeV contributes only a tiny fraction. 
It can be modelled by the respective pQCD result~\cite{Chetyrkin:1997pn} assigning a model uncertainty of 10\% to the cross section without introducing a numerically relevant error to the experimental moments. 
We emphasize that this continuum contribution should be regarded as a rough estimate for the (missing) experimental data rather than an additional theory input.
The precise numbers for the relevant experimental moments together with their statistical and systematical errors can be found in Ref.~\cite{Hoang:2012us}.

\section{Theoretical Moments at NNLL}
\label{thmoments}

For details on the derivation of the theoretical moments according to the scheme in Eq.~\eqref{counting} we refer to Refs.~\cite{Voloshin:1995sf,Hoang:1998uv,Hoang:1999ye,Hoang:2012us}. The resummation of nonrelativistic logarithms follows the vNRQCD approach~\cite{Hoang:2012us}. 
Here we shall only discuss the general structure and the latest computational progress concerning the RGI of the theoretical large-$n$ moments.
The theory prediction for the normalized total $b{\bar b}$ pair production cross section $R_{b\bar b}(s)$ in the threshold region is due to the optical theorem related to nonrelativistic current-current correlators, which describe the production and annihilation of a heavy quark pair. Explicit results for these correlators can be adopted from previous works on $t \,{\bar t}$ threshold production in $e^+e^-$ collisions~\cite{Hoang:2000ib,Hoang:2001mm,Hoang:2002yy,Hoang:2003ns}. 
After the moment integration over $s$ the result for the $n$-th moment through NNLL order can be expressed as
\begin{align}
P_n^{th,\rm NNLL} &=
\frac{3\,N_c\,Q_b^2\,\sqrt{\pi}}{4^{n+1} (M_b^{\rm pole})^{2n} \, n^{3/2}}\,
\bigg\{\,
c_1(h,\nu)^2\,\varrho_{n,1}(h,\nu) +
2\, c_1(h,\nu) c_2(h,\nu) \,\varrho_{n,2}(h,\nu)
\,\bigg\}\,,
\label{Pnth}
\end{align}
where the $\varrho_{n,i}$ arise from the integration of the nonrelativistic current correlators and the $c_i$ are Wilson coefficients of the respective effective currents. The variables $h$ and $\nu$ are introduced to parametrize the matching and renormalization scales of the effective theory. The natural choice is $h\sim 1$, $\nu\sim 1/\sqrt{n}$. The residual dependence of the bottom mass fit on these parameters is used for the perturbative error estimate in Sec.~\ref{results}. 

Equation~\eqref{Pnth} explicitly depends on the bottom pole mass $M_b^{\rm pole}$, which we translate to the 1S mass $M_b^{1S}$ using the relation
\begin{align}
M_b^{\rm pole} &= M_b^{1S} \{1+\Delta^{\rm LL} + \Delta^{\rm NLL} + [(\Delta^{\rm LL})^2 + \Delta_c^{\rm NNLL} + \Delta_m^{\rm NNLL}]\}\,.
\label{mpoletom1s}
\end{align}
The $\Delta$ terms are labeled according to the nonrelativistic order counting scheme in Eq.~\eqref{counting}. Explicit expressions can be found in Ref.~\cite{Hoang:2001mm}.
For the final theoretical expression used in the single moment fits below we consistently expand out the perturbative series for the Wilson coefficients $c_i$ together with the nonrelativistic expansion series for the $\varrho_{n,i}$ and $M_b^{\rm pole}$ in Eq.~\eqref{Pnth}.\footnote{We are forced to simultaneously expand out the series for $M_b^{\rm pole}$ and the $\varrho_{n,i}$ in the way explained in Ref.~\cite{Hoang:1999ye} in order to achieve a proper cancellation of the leading renormalon.}
The convergence properties of this expansion are discussed in detail in Ref.~\cite{Hoang:2012us}.

Apart from the NNLL correction to the renormalization group (RG) running of the Wilson coefficient $c_1$ associated with the dominant heavy quark production current all relevant contributions to Eq.~\eqref{Pnth} are known completely. Concerning the NNLL running of $c_1$ all (``non-mixing'') contributions from genuine vNRQCD three-loop diagrams were computed in Ref.~\cite{Hoang:2003ns}. The corresponding pNRQCD calculation is not available at present. The remaining NNLL (``mixing'') contributions are generated by corrections to the vNRQCD four-quark operator (``potential'') coefficients appearing in the NLL anomalous dimension of $c_1$. Recent results for the ultrasoft NLL running of the subleading ($\ord(v)$ and $\ord(v^2)$) nonrelativistic quark-antiquark potentials~\cite{Pineda:2011aw,Hoang:2011gy,Hoang:2006ht} completed the calculation of the ultrasoft part of the NNLL mixing contributions~\cite{Hoang:2011gy}. It is the dominant NNLL mixing effect~\cite{Hoang:2011it,Hoang:2012us} in the running of $c_1$. Likewise the ultrasoft terms dominate the NNLL non-mixing running~\cite{Hoang:2003ns}.
Thus at present the only unknown piece in $P_n^{th,\rm NNLL}$ is the NNLL soft mixing contribution to the RG evolution of $c_1$.

\begin{figure}[t]
\begin{center}
\includegraphics[width=0.55\textwidth]{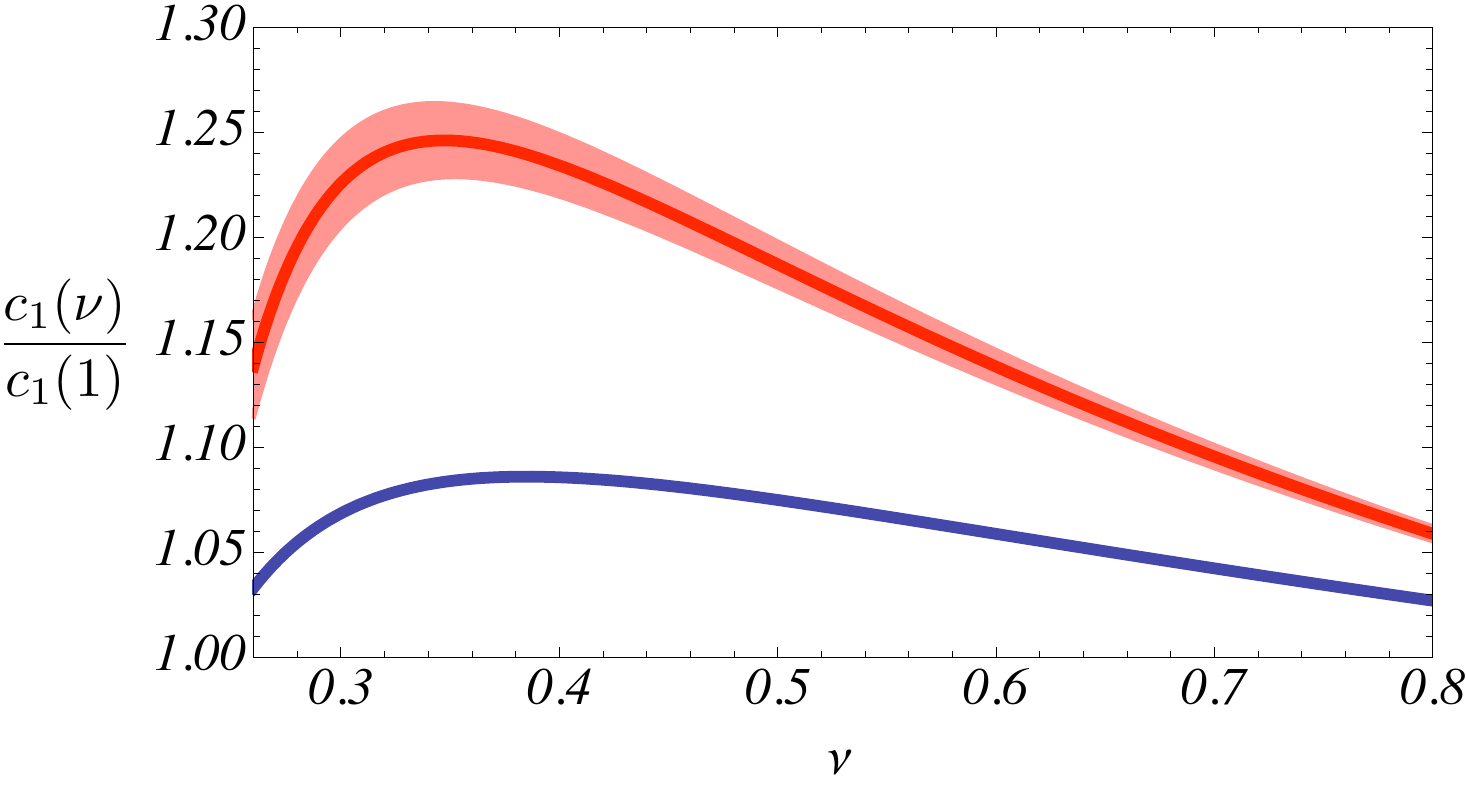}
\caption{RG evolution of the current coefficient $c_1$: NLL (blue) and approximate NNLL result (red) with uncertainty due to the unknown NNLL soft mixing contribution (light red band).\label{c1running}}
 \end{center}
\end{figure}

Figure~\ref{c1running} shows the dependence of the current coefficient $c_1(\nu)\equiv c_1(h\!=\!1,\nu)$ on the renormalization parameter $\nu$ for the complete NLL result (blue) and an approximate NNLL result (red), where all known NNLL contributions\footnote{These even include the first logarithm $\propto \alpha_s^3 \ln \nu$ in the NNLL series of the soft mixing contribution~\cite{Hoang:2003ns}. The subset of spin-dependent terms in the NNLL soft mixing contribution is also known~\cite{Penin:2004ay}, but tiny and neglected here. All relevant analytic NNLL expressions for $c_1$ are given in Ref.~\cite{Hoang:2012us}.} are added. The (light red) band around the NNLL curve is generated by varying all known soft NNLL contributions to that curve by a factor between 0 and 2.  
The uncertainty due to the unknown NNLL soft mixing terms estimated by this band is much smaller than the large total NNLL correction from the running in the relevant range $0.3\lesssim\nu\lesssim 0.6$ and can safely be neglected in the following~\cite{Hoang:2012us}. 

The (known) leading nonperturbative power correction to the moment $P_n^{th,\rm NNLL}$ is associated with the gluon condensate~\cite{Voloshin:1995sf} and turns out to be completely negligible for our analysis~\cite{Hoang:2012us}. Higher order power corrections are sufficiently suppressed for $n \lesssim 10$.

\section{Single Moment Fits}
\label{results}

\begin{figure}[t]
\begin{center}
\includegraphics[width=0.36\textwidth]{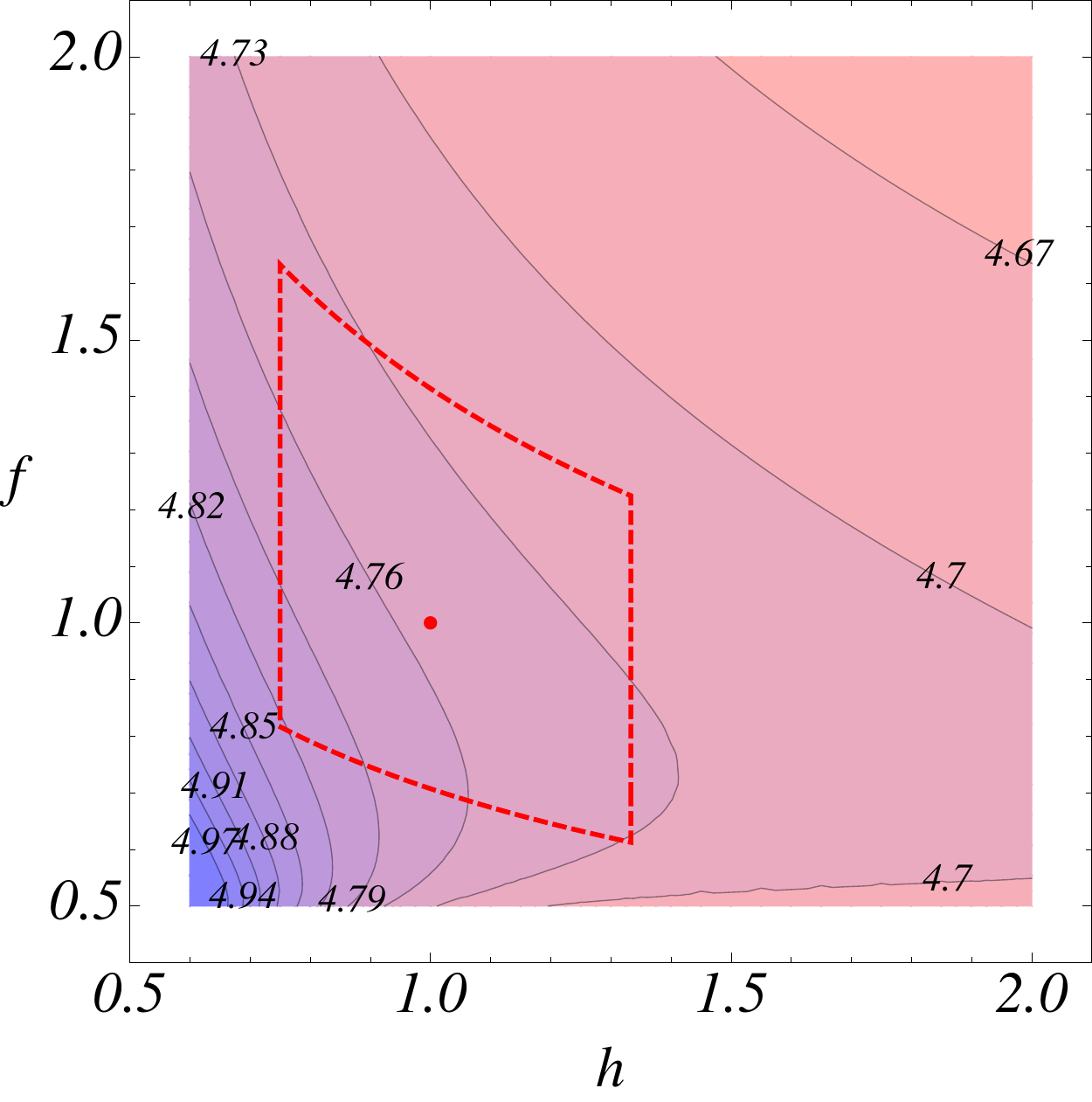}
\qquad
\includegraphics[width=0.5\textwidth]{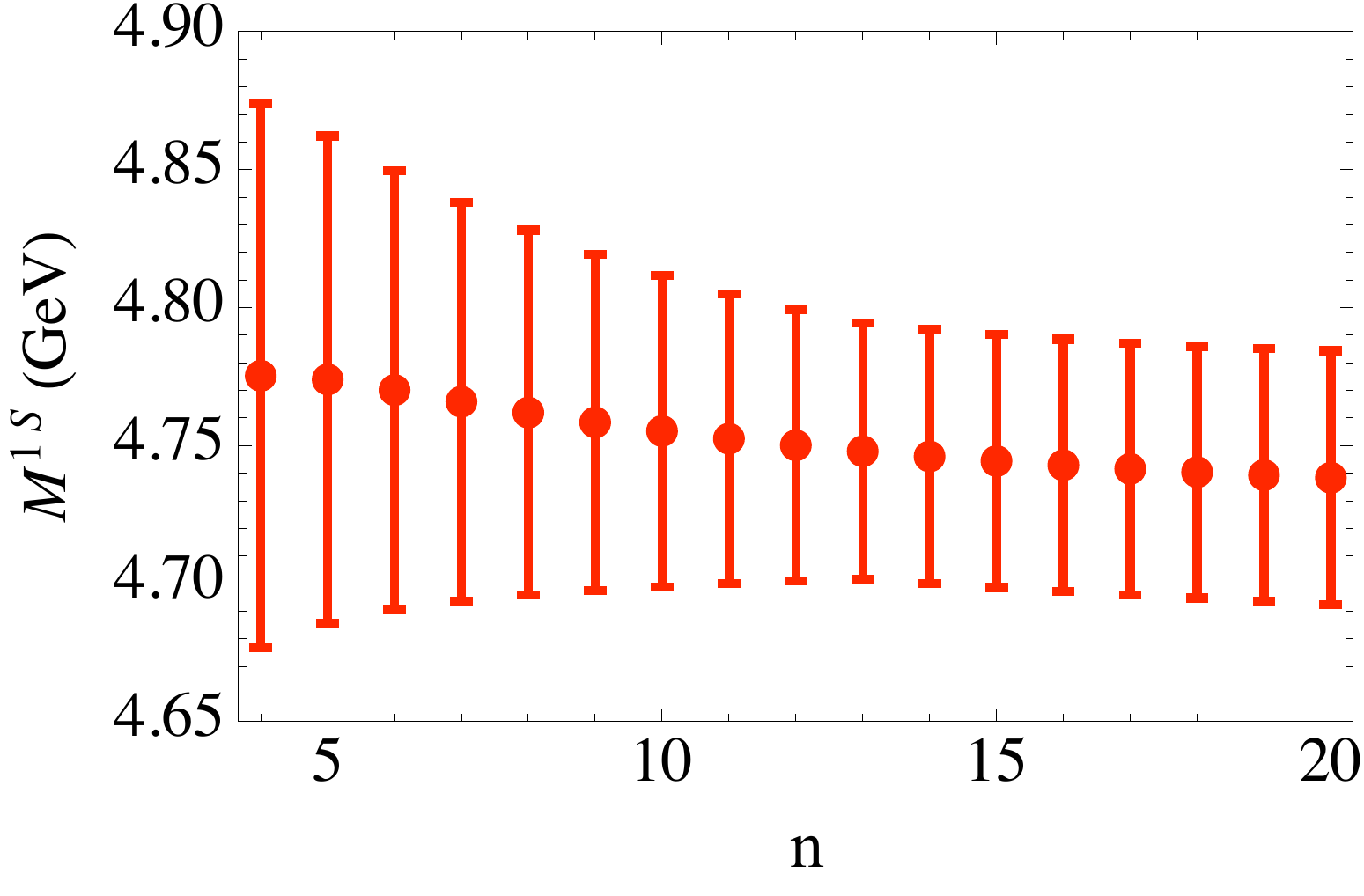}
\put(-410,148){a)}
\put(-215,148){b)}
\end{center}
\caption{
Panel a):
Contour plot of the 1S bottom mass determined from $P^{th}_{10}(m_b) =
  P^{exp}_{10}$ as a function of the parameters $h$ and $f\equiv \nu/\nu_*$. The different contours are labeled by the respective mass value in GeV. 
The region in the $h$-$f$ plane bounded by the red dashed line represents
the parameter space we scan to determine the variation of the mass, which
contributes to our perturbative error estimate. The region is defined by $0.75 \le h \le
1/0.75$ and demanding that $0.5\, \mu_U^* \le \mu_U \le 2\, \mu_U^*$, where
$\mu_U^*=m_b \nu_*^2$. 
The red point inside this area indicates our default values $f=h=1$ for the mass
determination.
Panel b): 1S mass results (dots) with perturbative error bars from single moment fits for $n=4$ to $n=20$ as explained in the text.
\label{ContourAndn420Plot}
} 
\end{figure}

The vNRQCD expression for the theoretical moment $P_n^{th,\rm NNLL}$ exhibits a residual dependence on the scale $\mu_h = h\, m_b$, where the effective theory is matched to full QCD, as well as on the two correlated renormalization scales $\mu_S = h\, m_b \nu$ (soft) and $\mu_U = h\, m_b \nu^2$ (ultrasoft). Here and in the following $m_b \equiv M_b^{1S}$. The three unphysical scales can be consistently parametrized by the two variables $h$ and $\nu$. In order to estimate the uncertainties from higher order perturbative corrections we choose to vary the parameters for the bottom mass fits within the $h$-$\nu$ region around the default values $\nu=\nu_*:=1/\sqrt{n} + 0.2$ and $h=1$ as defined in Fig.~\ref{ContourAndn420Plot}~a. The plot also shows the contours of the result for $M_b^{1S}$ from the equation $P_{10}^{th}(M_b^{1S}) = P_{10}^{exp}$, i.e. a NNLL single moment fit for $n=10$.

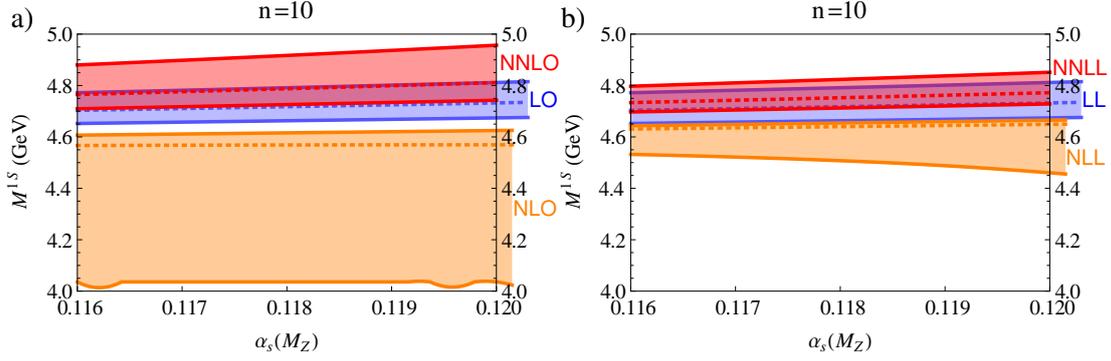
\begin{figure}[h]
\includegraphics[width=\textwidth]{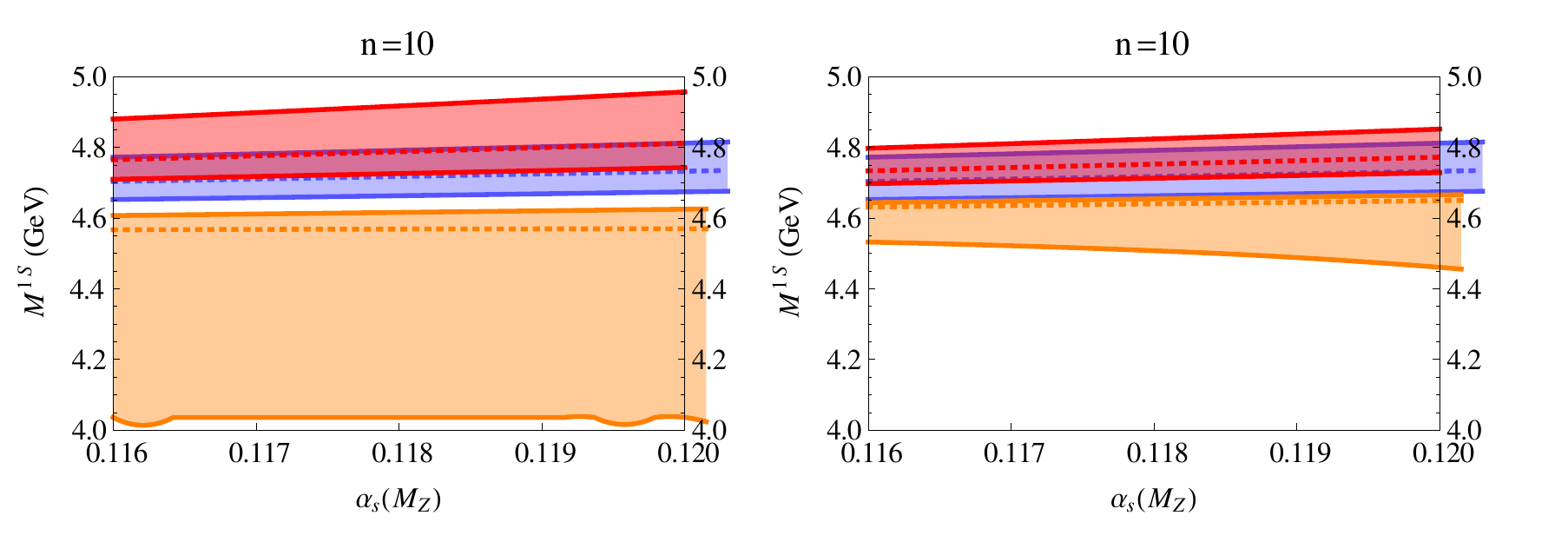}
\put(-230,100){\blau \sf \scriptsize LO}
\put(-236,59){\orange \sf \scriptsize NLO}
\put(-241,114){\rot \sf \scriptsize NNLO}
\put(-23,100){\blau \sf \scriptsize LL}
\put(-29,78){\orange \sf \scriptsize NLL}
\put(-34,112){\rot \sf \scriptsize NNLL}
\put(-424,130){a)}
\put(-218,130){b)}
\caption[]{Comparison of the masses obtained from the fixed order (a) and RGI calculation (b) of the 10-th moment, $P^{th}_{10}(m_b) = P^{exp}_{10}$. In panels a and b we show the  mass values with LO, NLO, NNLO and LL, NLL, NNLL accuracy, respectively. The dashed lines display the results from the fits with the default values for the parameters $h$ and $\nu$. The corresponding (partly overlapping) error bands were generated by varying $h$ and $\nu$ within the parameter space defined in
 Fig.~\ref{ContourAndn420Plot}~a. 
(We also added the tiny experimental error in quadrature, which is however hardly visible.) Concerning panel a, we note that for some low $m_b$ values in the NLO band and the associated values for $h$ and $\nu$ the ultrasoft coupling $\alpha_S(\mu_U)$ reaches $0.65$ causing numerical instabilities.
\label{RGIvsFixedOrder}
} 
\end{figure}

With these conventions for the $h$-$\nu$ scaling variations we can now generate error bands around the default fits (dashed lines) for $M_b^{1S}(\alpha_s)$ as shown in Fig.~\ref{RGIvsFixedOrder}.
The two panels in this figure compare the fits of the bottom mass as a function of the strong coupling $\alpha_s(M_Z)$ using theoretical moments calculated in the fixed-order (a) and the RGI approach (b). The fixed-order moments are obtained by switching off the all-order resummation of nonrelativistic logarithms in our RGI moments as explained in Ref.~\cite{Hoang:2012us}. Since at leading order (LO) the only relevant physical scale is the soft scale the LO and LL bands in Fig.~\ref{RGIvsFixedOrder} agree exactly. Comparing the next-to-leading (NLO) with the NLL and in particular the NNLO with the NNLL results we however observe much larger scale variations of the fixed-order results. This clearly indicates a substantially improved precision related to the resummation of logarithms in the RGI approach.
As argued in Ref.~\cite{Hoang:2012us} we believe that in contrast to the LL band, which is generated only by soft scale variations, and the (w.r.t. the default fits) strongly asymmetric NLL error bands the NNLL mass range gives a reliable estimate of the perturbative uncertainty. 
The observed bottom mass dependence on the input value for $\alpha_s(M_Z)$ is rather mild and at least in the interval $0.113\le\alpha_s(M_Z)\le 0.120$ linear to a good approximation~\cite{Hoang:2012us}.
A far more detailed analysis of the numerical results including plots of Fig.~\ref{RGIvsFixedOrder}~b type for different values of $n$ as well as multiple moment fits has been carried out in Ref.~\cite{Hoang:2012us}.

The final result for the 1S bottom mass from the NNLL RGI single moment analysis for $n=10$ outlined above is
\begin{align}
M_b^{1S} \, = \, 4.755\, \pm \, 0.057_{\rm pert}\, \pm\, 0.009_{\alpha_s}\,
\pm \, 0.003_{\rm exp}
\,\,\mbox{GeV}
\,,
\label{mbfinal1}
\end{align}
where we have used the current world average $\alpha_s(M_Z) = 0.1183 \pm 0.0010$ for the strong coupling~\cite{Bethke:2011tr}.
The central value in Eq.~\eqref{mbfinal1} corresponds to the result of the default fit ($h=1$, $\nu=\nu_*$). The quoted errors refer to the perturbative uncertainty, which we estimate by half the size of the band from the scale variations, and the errors from the uncertainties of $\alpha_s$ and the experimental data used for the fit.

Figure~\ref{ContourAndn420Plot}~b compares the $M_b^{1S}$ results and their respective perturbative error bars from fits using the moments $n=4$ to $n=20$. Within the errors all central values are in very good agreement. We however emphasize that for $n$ considerably larger than 10 formally the OPE for the theoretical moments breaks down due to possibly uncontrolled higher order power corrections, although the leading power correction still appears to be small as long as $n \lesssim 20$. We therefore regard the error bars for $n>10$ shown in Fig.~\ref{ContourAndn420Plot}~b as a confirmation of the perturbative stability of the RGI vNRQCD calculation, but do not use them for quoting our final errors. On the other hand the error bars for $n<10$ increase for smaller $n$ because the sensitivity of the theoretical moments on the mass decreases~\cite{Hoang:2012us}, cf. Eq.~\eqref{Pnth}.

Using the respective (fixed-order) relations to the pole mass, see Refs.~\cite{Hoang:1999ye,Hoang:2000fm} for details, we can translate the 1S mass result in Eq.~\eqref{mbfinal1} to the $\msb$-mass and obtain
\begin{align}
\overline m_b(\overline m_b) \, = \, 4.235\, \pm \, 0.055_{\rm pert}\,
\pm \, 0.003_{\rm exp}
\,\,\mbox{GeV}
\,,
\label{msbfinal1}
\end{align} 
where we have added an additional conversion error of $15$~MeV to the perturbative uncertainty~\cite{Hoang:2012us}.
Interestingly the $\alpha_s$ dependence of the original 1S mass result in Eq.~\eqref{mbfinal1} and the intrinsic $\alpha_s$ dependence of the 1S-$\msb$ conversion formula almost cancel exactly. The remaining $\alpha_s$ induced error is therefore negligible and not quoted in Eq.~\eqref{msbfinal1}.

\section{Summary}

We have presented the determination of the 1S bottom mass from (single) large-$n$ moment fits with RGI at NNLL order as carried out in Ref.~\cite{Hoang:2012us}. 
The main result is given in Eq.~\eqref{mbfinal1}. Converted to the $\msb$ scheme our result (Eq.~\eqref{msbfinal1}) is consistent with the NLL RGI large-$n$ result of Ref.~\cite{Pineda:2006gx}, but not quite compatible with the latest results from low-$n$ sum rule determinations~\cite{Chetyrkin:2009fv,Bodenstein:2011fv}. We however note that our calculation (like Ref.~\cite{Pineda:2006gx}) treats the charm quark as massless, while previous
fixed-order analyses~\cite{Hoang:1999us,Hoang:2000fm} have shown that finite charm mass effects are enhanced for large $n$ and cause a sizable mass shift between $-20$ and $-30$~MeV. A similar effect is expected in the RGI analysis and might help to reconcile the discrepancy.


\acknowledgments{
I would like to thank Andr\'e~Hoang for comments on the manuscript. This work was supported by the DFG under Emmy-Noether Grant No. TA 867/1-1.
}

\bibliographystyle{JHEP-2NoTitle}
\bibliography{MeiBmassBib2}

\providecommand{\href}[2]{#2}\begingroup\raggedright\begin{thebibliography}{10}

\bibitem{Beringer:1900zz}
J.~Beringer {\em et.~al.} (Particle Data Group) {\em
  Phys.Rev.} {\bf D86} (2012) 010001.

\bibitem{Novikov:1977dq}
V.~Novikov, L.~Okun, M.~A. Shifman, A.~Vainshtein, M.~Voloshin {\em et.~al.}
  {\em Phys.Rept.} {\bf 41} (1978) 1--133.

\bibitem{Chetyrkin:2009fv}
K.~Chetyrkin, J.~Kuhn, A.~Maier, P.~Maierhofer, P.~Marquard {\em et.~al.} {\em
  Phys.Rev.} {\bf D80} (2009) 074010
  [\href{http://arXiv.org/abs/0907.2110}{{\tt 0907.2110}}].

\bibitem{Chetyrkin:2010ic}
K.~Chetyrkin, J.~Kuhn, A.~Maier, P.~Maierhofer, P.~Marquard {\em et.~al.} {\em
  Theor.Math.Phys.} {\bf 170} (2012) 217--228
  [\href{http://arXiv.org/abs/1010.6157}{{\tt 1010.6157}}].

\bibitem{Bodenstein:2011fv}
S.~Bodenstein, J.~Bordes, C.~Dominguez, J.~Penarrocha and K.~Schilcher {\em
  Phys.Rev.} {\bf D85} (2012) 034003
  [\href{http://arXiv.org/abs/1111.5742}{{\tt 1111.5742}}].

\bibitem{Corcella:2002uu}
G.~Corcella and A.~Hoang {\em Phys.Lett.} {\bf B554} (2003) 133--140
  [\href{http://arXiv.org/abs/hep-ph/0212297}{{\tt hep-ph/0212297}}].

\bibitem{Caswell:1985ui}
W.~Caswell and G.~Lepage {\em Phys.Lett.} {\bf B167} (1986) 437.

\bibitem{Bodwin:1992ye}
G.~T. Bodwin, E.~Braaten and G.~P. Lepage {\em Phys.Rev.} {\bf D46} (1992)
  1914--1918 [\href{http://arXiv.org/abs/hep-lat/9205006}{{\tt
  hep-lat/9205006}}].

\bibitem{pNRQCDfirst}
A.~Pineda and J.~Soto {\em Nucl.Phys.Proc.Suppl.} {\bf 64} (1998) 428--432
  [\href{http://arXiv.org/abs/hep-ph/9707481}{{\tt hep-ph/9707481}}].

\bibitem{Brambilla:1999xf}
N.~Brambilla, A.~Pineda, J.~Soto and A.~Vairo {\em Nucl.Phys.} {\bf B566}
  (2000) 275 [\href{http://arXiv.org/abs/hep-ph/9907240}{{\tt
  hep-ph/9907240}}].

\bibitem{Luke:1999kz}
M.~E. Luke, A.~V. Manohar and I.~Z. Rothstein {\em Phys.Rev.} {\bf D61} (2000)
  074025 [\href{http://arXiv.org/abs/hep-ph/9910209}{{\tt hep-ph/9910209}}].

\bibitem{Pineda:2006gx}
A.~Pineda and A.~Signer {\em Phys.Rev.} {\bf D73} (2006) 111501
  [\href{http://arXiv.org/abs/hep-ph/0601185}{{\tt hep-ph/0601185}}].

\bibitem{Pineda:2006ri}
A.~Pineda and A.~Signer {\em Nucl.Phys.} {\bf B762} (2007) 67--94
  [\href{http://arXiv.org/abs/hep-ph/0607239}{{\tt hep-ph/0607239}}].

\bibitem{Voloshin:1995sf}
M.~Voloshin {\em Int.J.Mod.Phys.} {\bf A10} (1995) 2865--2880
  [\href{http://arXiv.org/abs/hep-ph/9502224}{{\tt hep-ph/9502224}}].

\bibitem{Hoang:1998uv}
A.~Hoang {\em Phys.Rev.} {\bf D59} (1999) 014039
  [\href{http://arXiv.org/abs/hep-ph/9803454}{{\tt hep-ph/9803454}}].

\bibitem{Melnikov:1998ug}
K.~Melnikov and A.~Yelkhovsky {\em Phys.Rev.} {\bf D59} (1999) 114009
  [\href{http://arXiv.org/abs/hep-ph/9805270}{{\tt hep-ph/9805270}}].

\bibitem{Beneke:1999fe}
M.~Beneke and A.~Signer {\em Phys.Lett.} {\bf B471} (1999) 233--243
  [\href{http://arXiv.org/abs/hep-ph/9906475}{{\tt hep-ph/9906475}}].

\bibitem{Hoang:1999ye}
A.~Hoang {\em Phys.Rev.} {\bf D61} (2000) 034005
  [\href{http://arXiv.org/abs/hep-ph/9905550}{{\tt hep-ph/9905550}}].

\bibitem{Hoang:2000fm}
A.~Hoang \href{http://arXiv.org/abs/hep-ph/0008102}{{\tt hep-ph/0008102}}.

\bibitem{Battaglia:2003in}
M.~Battaglia, A.~Buras, P.~Gambino, A.~Stocchi, D.~Abbaneo {\em et.~al.}
  \href{http://arXiv.org/abs/hep-ph/0304132}{{\tt hep-ph/0304132}}.

\bibitem{Beneke:1998rk}
M.~Beneke {\em Phys.Lett.} {\bf B434} (1998) 115--125
  [\href{http://arXiv.org/abs/hep-ph/9804241}{{\tt hep-ph/9804241}}].

\bibitem{Hoang:1999zc}
A.~Hoang and T.~Teubner {\em Phys.Rev.} {\bf D60} (1999) 114027
  [\href{http://arXiv.org/abs/hep-ph/9904468}{{\tt hep-ph/9904468}}].

\bibitem{Hoang:2000yr}
A.~Hoang, M.~Beneke, K.~Melnikov, T.~Nagano, A.~Ota {\em et.~al.} {\em
  Eur.Phys.J.direct} {\bf C2} (2000) 1
  [\href{http://arXiv.org/abs/hep-ph/0001286}{{\tt hep-ph/0001286}}].

\bibitem{Pineda:2001zq}
A.~Pineda {\em JHEP} {\bf 0106} (2001) 022
  [\href{http://arXiv.org/abs/hep-ph/0105008}{{\tt hep-ph/0105008}}].

\bibitem{Hoang:2012us}
A.~Hoang, P.~Ruiz-Femenia and M.~Stahlhofen {\em JHEP} {\bf 1210} (2012) 188
  [\href{http://arXiv.org/abs/1209.0450}{{\tt 1209.0450}}].

\bibitem{Aubert:2008ab}
B.~Aubert {\em et.~al.} (BABAR Collaboration) {\em
  Phys.Rev.Lett.} {\bf 102} (2009) 012001
  [\href{http://arXiv.org/abs/0809.4120}{{\tt 0809.4120}}].

\bibitem{Dehnadi:2011gc}
B.~Dehnadi, A.~H. Hoang, V.~Mateu and S.~M. Zebarjad
  \href{http://arXiv.org/abs/1102.2264}{{\tt 1102.2264}}.

\bibitem{Chetyrkin:1997pn}
K.~Chetyrkin, A.~Hoang, J.~H. Kuhn, M.~Steinhauser and T.~Teubner {\em
  Eur.Phys.J.} {\bf C2} (1998) 137--150
  [\href{http://arXiv.org/abs/hep-ph/9711327}{{\tt hep-ph/9711327}}].

\bibitem{Hoang:2000ib}
A.~Hoang, A.~Manohar, I.~W. Stewart and T.~Teubner {\em Phys.Rev.Lett.} {\bf
  86} (2001) 1951--1954 [\href{http://arXiv.org/abs/hep-ph/0011254}{{\tt
  hep-ph/0011254}}].

\bibitem{Hoang:2001mm}
A.~Hoang, A.~Manohar, I.~W. Stewart and T.~Teubner {\em Phys.Rev.} {\bf D65}
  (2002) 014014 [\href{http://arXiv.org/abs/hep-ph/0107144}{{\tt
  hep-ph/0107144}}].

\bibitem{Hoang:2002yy}
A.~H. Hoang and I.~W. Stewart {\em Phys.Rev.} {\bf D67} (2003) 114020
  [\href{http://arXiv.org/abs/hep-ph/0209340}{{\tt hep-ph/0209340}}].

\bibitem{Hoang:2003ns}
A.~H. Hoang {\em Phys.Rev.} {\bf D69} (2004) 034009
  [\href{http://arXiv.org/abs/hep-ph/0307376}{{\tt hep-ph/0307376}}].

\bibitem{Pineda:2011aw}
A.~Pineda {\em Phys.Rev.} {\bf D84} (2011) 014012
  [\href{http://arXiv.org/abs/1101.3269}{{\tt 1101.3269}}].

\bibitem{Hoang:2011gy}
A.~H. Hoang and M.~Stahlhofen {\em JHEP} {\bf 1106} (2011) 088
  [\href{http://arXiv.org/abs/1102.0269}{{\tt 1102.0269}}].

\bibitem{Hoang:2006ht}
A.~H. Hoang and M.~Stahlhofen {\em Phys.Rev.} {\bf D75} (2007) 054025
  [\href{http://arXiv.org/abs/hep-ph/0611292}{{\tt hep-ph/0611292}}].

\bibitem{Hoang:2011it}
A.~Hoang and M.~Stahlhofen \href{http://arXiv.org/abs/1111.4486}{{\tt
  1111.4486}}.

\bibitem{Penin:2004ay}
A.~A. Penin, A.~Pineda, V.~A. Smirnov and M.~Steinhauser {\em Nucl. Phys.} {\bf
  B699} (2004) 183--206 [\href{http://arXiv.org/abs/hep-ph/0406175}{{\tt
  hep-ph/0406175}}].

\bibitem{Bethke:2011tr}
S.~Bethke, A.~H. Hoang, S.~Kluth, J.~Schieck, I.~W. Stewart {\em et.~al.}
  \href{http://arXiv.org/abs/1110.0016}{{\tt 1110.0016}}.

\bibitem{Hoang:1999us}
A.~Hoang and A.~Manohar {\em Phys.Lett.} {\bf B483} (2000) 94--98
  [\href{http://arXiv.org/abs/hep-ph/9911461}{{\tt hep-ph/9911461}}].

\end{thebibliography}\endgroup

\end{document}